\documentclass[10pt,journal]{IEEEtran}
\usepackage[utf8]{inputenc}
\usepackage{textcomp}
\usepackage{caption}

\usepackage[perpage,symbol]{footmisc}
\usepackage[english]{babel}
\usepackage{multirow,multicol}
\usepackage{amsmath, bm} 
\usepackage{pbox, booktabs, rotating}
\usepackage{array, makecell, lipsum}
\usepackage{enumerate}
\usepackage{threeparttable}
\usepackage{multirow,hyphenat,balance}
\usepackage{placeins}
\usepackage[compress]{cite}
\usepackage{float,color}
\usepackage{ragged2e}
\usepackage{enumerate}
\usepackage{etoolbox}
\usepackage{paralist}
\usepackage{cases}
\usepackage{url}
\usepackage{framed}
\usepackage{flushend}
\usepackage{amsmath}
\usepackage{amsthm}
\usepackage{amsthm}
\usepackage[english]{babel}
\usepackage{amssymb}
\usepackage[linesnumbered,ruled]{algorithm2e}
\let\oldnl\nl% Store \nl in \oldnl
\newcommand{\nonl}{\renewcommand{\nl}{\let\nl\oldnl}}% Remove line number for one line
\usepackage{pifont}
\newcommand{\cmark}{\ding{51}}%
\newcommand{\xmark}{\ding{55}}%
\usepackage{commath}

\newcommand*{\eg}{\textit{e.g.,}\@\xspace}
\newcommand*{\el}{et al.\@\xspace}

\newcolumntype{M}[1]{>{\centering\arraybackslash}m{#1}}

\DeclareCaptionLabelFormat{nospace}{#1#2}
\newcolumntype{"}{@{\hskip\tabcolsep\vrule width 1.5pt\hskip\tabcolsep}}
\makeatother

\author{Yadi Zhong, \IEEEmembership{Student Member IEEE} and Ujjwal Guin, \IEEEmembership{Senior Member IEEE}

\thanks{Yadi Zhong and Ujjwal Guin are with the Department of Electrical and Computer Engineering, Auburn University, AL, USA (e-mail: \{yadi and ujjwal.guin\}@auburn.edu).}

}

\title{A Comprehensive Test Pattern Generation Approach Exploiting the SAT Attack for Logic Locking} %may be different

\begin{document}

\maketitle

\begin{abstract}
The need for reducing manufacturing defect escape in today's safety-critical applications requires increased fault coverage. However, generating a test set using commercial automatic test pattern generation (ATPG) tools that lead to zero-defect escape is still an open problem. It is challenging to detect all stuck-at faults to reach 100\% fault coverage. In parallel, the hardware security community has been actively involved in developing solutions for logic locking to prevent IP piracy. In logic locking, locks are inserted in different locations of the netlist to modify the original functionality. Unless the correct key is programmed into the IC, the circuit functions incorrectly. Unfortunately, the Boolean satisfiability (SAT) based attack, introduced in~\cite{subramanyan2015evaluating}, can determine the secret key efficiently, and break different logic locking schemes. In this paper, we propose a novel test pattern generation approach using the powerful SAT attack on logic locking. A stuck-at fault is modeled as a locked gate with a secret key, where it can effectively deduce the satisfiable assignment with reduced backtracks under key initialization of the SAT attack. The input pattern that determines the key is a test for the stuck-at fault. We propose two different approaches for test pattern generation. First, a single stuck-at fault is targeted, and a corresponding locked circuit with one key bit is created. This approach generates one test pattern per fault. Second, we consider a group of faults and convert the circuit to its locked version with multiple key bits. The inputs obtained from the SAT attack tool are the test set for detecting this group of faults. Our approach can find test patterns for all hard-to-detect faults that were previously undetected in commercial ATPG tools. The proposed test pattern generation approach can efficiently detect redundant faults as well. We demonstrate the effectiveness of the approach on ITC'99 benchmarks. The results show that we can detect all the hard-to-detect faults and identify redundant faults and a 100\% stuck fault coverage is achieved. In addition, we show that test generation time saving becomes significant for Approach 2 as multiple faults help reduce or remove conflicts.
\end{abstract}

\vspace{5px}

\section{Introduction} \label{sec:intro}
\IEEEPARstart{T}{he} exponential growth of integrated circuits (ICs) in our critical infrastructure requires aggressive testing as system failure has severe safety consequences. As a result, it is critical that the escape of manufacturing defects to the next stage approaches zero. For example, multiple safety standards, like AEC-Q100 and ISO 26262~\cite{iso_2018}, are defined to meet the zero-defective-parts-per-million goal for safety-critical automotive chips. Testing plays a vital role in detecting all possible defects in the manufactured chips to avoid the potentially devastating effects when defective ones slip from the testing facility. Today's commercial automatic test pattern generation (ATPG) can generate test patterns for stuck-at, delay, bridging, and a few other fault models~\cite{SynopsysTetraMAX}. However, achieving a fault coverage that leads to zero-defect escape is still an open problem. For example, it is challenging to reach 100\% stuck-at fault coverage using commercial ATPG tools. It can be extremely difficult to sensitize a hard-to-detect fault and propagate the faulty response at the outputs of a large circuit, limiting the desired goal of achieving perfect fault coverage. It is also challenging to identify all the redundant faults as their effects cannot be propagated to the output. These faults can be ignored for determining a meaningful fault coverage, as they do not impact the function of a circuit. 

Over the past few decades, we have seen a steady increase in the fault coverage for digital circuits using the continued advancement in combinational ATPG techniques, from Roth's D-Algorithm~\cite{roth1966diagnosis,roth1967programmed} to PODEM~\cite{goel1981implicit}, FAN~\cite{fujiwara1983acceleration}, SOCRATES~\cite{schulz1988socrates}, TRAN~\cite{chakradhar1993transitive}, \textit{etc}. Along with these techniques, SAT-based test pattern generation has also been proposed as a solution to achieve higher fault coverage~\cite{larrabee1992test, stephan1996combinational, eggersgluss2007improving, eggersgluss2012new, eggersgluss2013improved, drechsler2008acceleration, fujita2014efficient, balcarek2013techniques}. For SAT-based techniques, the miter construction between the fault-free circuit and faulty circuit with a stuck-at fault (\textit{saf}) is the core for generating a test pattern for detecting that \textit{saf}. The increased number of conflicts for larger circuits resulting from the miter circuits and getting resolved at a later stage makes hard-to-detect faults undetectable. As these SAT-based prior works have been concluded nearly a decade ago, it is fair to assume that the Industry has already assimilated the state-of-the-art research. However, we still observe several hard-to-detect and redundant faults that a commercial ATPG tool, \eg Synopsys TetraMAX II~\cite{SynopsysTetraMAX} fails to identify even with the maximum abort limit (see Section~\ref{sec:result}). Some undetected faults are redundant faults in the circuit, where no pattern could propagate the faulty effect to the primary output. Others are the hard-to-detect faults, where the ATPG tools fall short in finding the appropriate test patterns even if such tests exist to detect these faults. Therefore, the main bottleneck from reaching high fault coverage for IC testing is in the undetected faults, specifically, the classification of redundant faults and test pattern generation for hard-to-detect faults. \textit{The focus of this paper is to analyze and classify these undetected faults, not identified by commercial ATPG tools, so that ($i$) we can accurately distinguish any redundant faults from non-redundant ones; ($ii$) generate the test patterns for each hard-to-detect faults; and ($iii$) generate tests for combining multiple hard-to-detect faults to reduce the total pattern count.}

\setlength{\textfloatsep}{5pt}
 \begin{figure}[t]
    \centering 
    \includegraphics[width=\columnwidth]{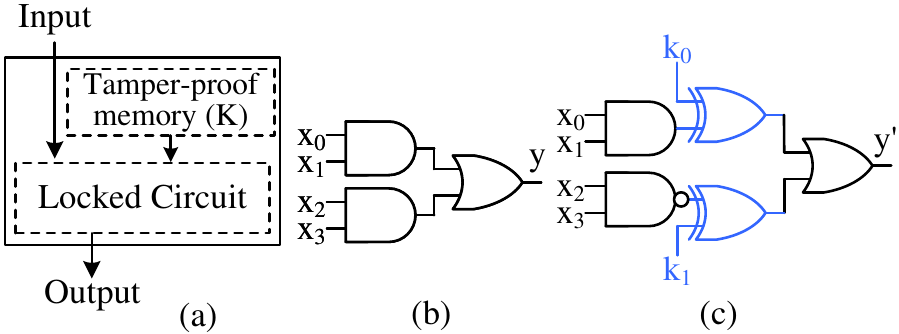} %\vspace{-15px}
    \caption{Logic Locking. (a) Overview of logic locking (b) Original circuit. (c) XOR-based locking with $\{k_0 k_1\}=\{01\}$.} \label{fig:logiclocking} 
    %  \vspace{-10px}
\end{figure}

The hardware security community has been actively involved in solving the threat of intellectual property (IP) piracy~\cite{castillo2007ipp, tehranipoor2011introduction, tehranipoor2015counterfeit, bhunia2018hardware, zhang2018chip} and IC overproduction~\cite{roy2008epic, alkabani2007active, chakraborty2008hardware, alkabani2007remote, huang2008ic, guin2016fortis, baumgarten2010preventing}, originating from the horizontal integration of semiconductor design, manufacturing, and test. It is practically infeasible for many design houses to manufacture chips on their own due to the increased chip design complexity and manufacturing processes. An untrusted entity in the semiconductor supply chain can pirate the design details and cause irreparable damage. Logic locking~\cite{roy2008epic, baumgarten2010preventing, rajendran2012security, guin2016fortis, rajendran2015fault} was proposed to counter IP piracy, where a circuit design is obfuscated using a secret key. Figure~\ref{fig:logiclocking} shows an abstract representation of logic locking with a simple example. The secret key ($K$) is programmed into the tamper-proof memory, as shown in Figure~\ref{fig:logiclocking}(a). Figure~\ref{fig:logiclocking}(b) shows an example of the original netlist with function $y=x_0x_1+x_2x_3$. A lock is inserted using two XOR gates with key $\{k_0 k_1\}$, as shown in Figure~\ref{fig:logiclocking}(c). This modifies the original functionality $y$ to $y'=(x_0x_1\oplus k_0)+(\overline{x_2x_3}\oplus k_1)$. The secret key value $\{k_0 k_1\}=\{01\}$ maps $y'$ to $y$ for all possible input combinations. The security of any locking scheme relies on the secrecy of the key. The original netlist can be recovered if an adversary obtains the correct key values. Subramanyan \el showed that Boolean Satisfiability (SAT) could be used to break traditional locking schemes effectively~\cite{subramanyan2015evaluating}. The attack constructs a miter circuit and asks SAT solver to find an input pattern that produces differentiating output behavior between incorrect keys and the right one, similar to revealing the faulty state to the output~\cite{rajendran2015fault, jain2020atpg, zhong2022afia}. The attack is very effective in determining the key (i.e., the value of $k$, the key-input of XOR shown in Figure~\ref{fig:logiclocking}(c)) no matter where the key gate (XOR) is placed in the netlist. This motivates us to develop a novel test pattern generation scheme using this powerful SAT attack. The question is, can we model a stuck-at fault to its key-dependent locked circuit counterpart so that the SAT attack can find a test pattern to determine the key, and thus a test for the same stuck-at fault?

In this paper, we show the novel miter construction for test pattern generation of stuck-at-0 (\textit{sa0}) and stuck-at-1 (\textit{sa1}) faults, where each fault has its equivalent locked circuit to be applied with the existing powerful SAT-based attack~\cite{subramanyan2015evaluating}. We target undetected faults where commercial ATPG tools~\cite{SynopsysTetraMAX} fall short in producing test patterns. Our work focuses on identifying the redundant ones from these undetected faults and finding suitable patterns for detecting non-redundant faults to increase fault coverage further. %A \textit{sa0} fault is modeled using an AND gate inserted into the fault site (see Figure~\ref{fig:sa0and}(a), (b)), whereas a \textit{sa1} fault is modeled using an OR gate (see Figure~\ref{fig:sa0and}(c), (d)). 
The equivalence of a stuck-at fault (\textit{sa0}, \textit{sa1}) is an AND or OR key-gate, respectively. Once the stuck fault is converted to a key-dependant AND/OR gate, we then ask the SAT attack~\cite{subramanyan2015evaluating} to solve the key and return the distinguishing input patterns it used in deriving the key value. A distinguishing input pattern returned by the SAT attack allows us to sensitize a stuck fault and propagate the faulty response to the output. \textit{To the best of our knowledge, this research is the first attempt to apply the SAT-based logic locking attack on the equivalent keyed circuit to ($i$) find test patterns for undetected faults, ($ii$) identify redundant faults, and ($iii$) reduce test pattern count for the combination of multiple faults.}

\vspace{5px}
The contributions of this paper are summarized as follows: 
\begin{itemize}
    % \item \textit{Modeling of stuck-at faults:} We propose a novel technique to model stuck-at faults (\eg \textit{sa0} and \textit{sa1}) to their equivalent locked circuits so that the SAT attack tool can be used to generate tests for faults that are undetected by the commercial tool. The novel fault modeling helps to reduce the number of backtracking during test pattern generation due to key initialization in the SAT attack. The time saving becomes significant when we consider multiple faults resulting in reducing or removing conflicts by several key bits.  
    \item \textit{Novel miter construction:} We propose a novel technique for test pattern generation using the proposed miter construction, which is extensively used for breaking logic locking. The stuck-at faults (\eg \textit{sa0} and \textit{sa1}) are modeled to their equivalent locked circuits so that the SAT attack tool can be used to generate tests for faults that are undetected by the commercial tool. The novel miter helps to reduce the number of backtracking during test pattern generation due to key initialization in the SAT attack. The time saving becomes significant when we consider multiple faults resulting in reducing or removing conflicts by several key bits.
    
    \vspace{3px}
    \item \textit{Test pattern generation for hard-to-detect faults:} 
    Upon the successful decryption of locking, the distinguishing input patterns (DIPs) returned from the SAT attack~\cite{subramanyan2015evaluating} are the desired test patterns. We propose two approaches for generating test patterns. The first approach considers one fault at a time, whereas the second approach combines multiple faults during test pattern generation. We believe we are the first to show that all the faults can either be detected or identified as redundant, and we can achieve a 100\% stuck fault coverage, including \textit{b19\_C} benchmark.
    
    \vspace{3px}
    \item \textit{Identification of redundant faults:} The proposed SAT-based test pattern generation approach can also identify any redundant faults. If a fault resides at a redundant site, the SAT program will not return any DIP. For any locked circuit with a redundant fault, unlocking it with the incorrect key does not change the circuit's functionality, where the same input-output pair is observed for both $k=0$ and $k=1$. As a result, no test coverage is necessary as the faulty response does not affect the output and can be discarded from the total detectable fault count. 
    
    %Our proposed approaches can identify redundant faults in a circuit. If a fault resides at a redundant site, the SAT program will not return any DIPs to differentiate $k=0$ and $k=1$, and thus it would reach the UNSAT conclusion.
    
\end{itemize}

The rest of the paper is organized as follows. We begin with a brief introduction to test pattern generation and the SAT attack on logic locking in Section~\ref{sec:background}. Our proposed approach to increase fault coverage is presented in Section~\ref{sec:proposed}. The result and analysis for the proposed approach are described in Section \ref{sec:result}. Finally, we conclude the paper in Section \ref{sec:conclusion}.

\section{Background} \label{sec:background}
Since our focus is to generalize any circuit with stuck-at faults to its equivalent key-dependent logic locking counterpart, we describe the working principle of both test pattern generation and the SAT attack~\cite{subramanyan2015evaluating} against logic locking while emphasizing the similarity between the two. We assume that the circuit of interest is purely combinational. Any sequential circuit is assumed to have scan chains, and we can perform scan-based testing so that it is analogous to a combinational circuit with additional pseudo-primary input (PPI) and pseudo-primary output (PPO) from the scan flip-flops~\cite{bushnell2004essentials}. 

\vspace{-5px}
\subsection{Test Pattern Generation} \label{subsec:background-ATPG}
Although various ATPG algorithms~\cite{roth1966diagnosis,goel1981implicit,fujiwara1983acceleration, kirkland1987topological, schulz1988socrates, giraldi1991est,chakradhar1993transitive, bushnell2004essentials} have been proposed over the past few decades, we outline Roth's D-Algorithm~\cite{roth1966diagnosis, roth1967programmed} in this section since it is the foundation for the subsequently revised ATPG techniques and the motivation for our proposed approach. D-algorithm consists of fault sensitization, fault propagation, and line justification~\cite{bushnell2004essentials}. %To generate a test pattern, ATPG first places a logic value (logic 1 or 0) opposite to the stuck-at fault (\textit{sa0} or \textit{sa1}) during fault sensitization. $D$ or $\overline{D}$ is assigned at the \textit{sa0} or \textit{sa1} locations, respectively, with D-algorithm. This is the fault sensitization necessary to differentiate a faulty circuit from a fault-free one. Then, ATPG performs fault propagation such that the fault response, either $D$ or $\overline{D}$, is observable at the primary output. Finally, in line justification, ATPG finds an input pattern in which fault sensitization and propagation are guaranteed. 
Conflict may occur during the logic assignment in both fault propagation and line justification, in which the ATPG has to backtrack to remove the previous assignment and make new decisions. What makes test pattern generation the NP-complete problem is that the actual number of backtracks and/or forward implication is agnostic to the ATPG tool, and the worst case is to iterate through all possible assignments forcing the complexity to become exponential to the circuit size~\cite{bushnell2004essentials}. To avoid the exponential running time when generating test patterns, the current ATPG tool~\cite{SynopsysTetraMAX} includes an upper limit to the number of possible backtracks before moving on with the subsequent fault.

In parallel, the research community explored SAT-based test pattern generation. The initial SAT-based technique, proposed by Larrabee~\cite{larrabee1992test}, constructs the Boolean difference between the faulty and fault-free circuits to detect single stuck-at faults. Stephan~\el proposed TEGUS that uses gate characteristic functions added in depth-first search order from inputs to outputs~\cite{stephan1996combinational}. Over the years, different SAT-based test pattern generation techniques have been proposed to obtain a high fault coverage~\cite{eggersgluss2007improving, drechsler2008acceleration,eggersgluss2012new,  balcarek2013techniques, eggersgluss2013improved, fujita2014efficient}. Eggersgluss~\el explored SAT-based test compaction with a large number of unspecified bits~\cite{eggersgluss2007improving, eggersgluss2012new, eggersgluss2013improved}. Drechsler~\el includes the modeling of tristate elements with additional unknown ($U$) and high-impedance ($Z$) states~\cite{drechsler2008acceleration}. Balcarek~\el~\cite{balcarek2013techniques} filters the unexcitable faults based on the static and dynamic implications. Fujita~\el\cite{fujita2014efficient} targets test pattern generation with multiple faults. However, to find a test pattern for a hard-to-detect fault, the SAT solver encounters a large number of conflicting assignments and requires an increased number of backtracks, which makes test generation time excessively high. 

\begin{figure}[ht]
    \centering  
    \includegraphics[width=\columnwidth]{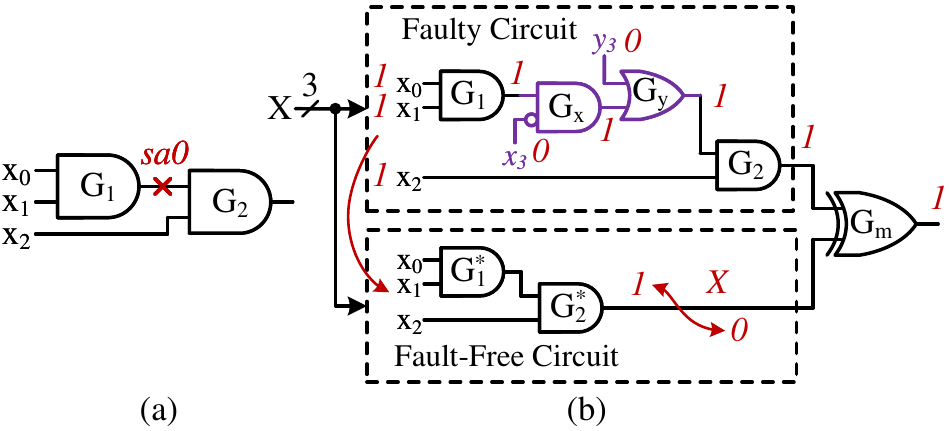}%\vspace{-5px}
    \caption{Conflicts in solving miter circuit presented in~\cite{fujita2014efficient}. (a) A simple circuit with a \textit{sa0} fault. (b) Conflict during SAT assignment.} \label{fig:fujita-miter} \vspace{-5px}
\end{figure}

Fujita~\el~\cite{fujita2014efficient} constructs the miter with a faulty circuit with modeled faults and fault-free circuits. Figure~\ref{fig:fujita-miter}(a) shows a simple circuit with a \textit{sa0} fault. Figure~\ref{fig:fujita-miter}(b) shows the miter circuit used for test generation~\cite{fujita2014efficient}. As the OR gate for modeling \textit{sa1} should remain ineffective as we are targeting \textit{sa0}, one can simply ignore it by having its input $y_3$ tied to $0$. Without loss of generality, we assume the SAT solver first makes a decision to the upper input node (corresponding to the output of the faulty circuit) of miter XOR gate $G_m$ with logic 1. With unit clause propagation, the output of the faulty circuit is 1, and all nodes inside the faulty circuit can be uniquely determined. Conflict arises at the fault-free circuit's Boolean assignment as the input $\{x_0,x_1,x_2\}=\{\text{111}\}$ derived from the faulty circuit cannot satisfies the required logic 0 output for the fault-free circuit. Therefore, the SAT solver is required to perform backtracks and resolves the conflict with logic 0 decision for the faulty circuit response and 1 for fault-free. In addition, if a conflict arises, it can be determined after logic assignments for all the nodes. This motivates us to construct a miter that reduces the possible backtracks or resolves conflicts at an earlier stage to ensure manageable test generation time.

\vspace{-5px}
\subsection{SAT attack on Logic Locking}\label{sec:satattack}
Over the years, the optimization and advancement of SAT algorithms have led to a significant decrease in average runtime for SAT solvers~\cite{fichte2020time}. This leads to a growing number of SAT-based applications. One of the most prominent attacks to counter logic locking, proposed by Subramanyan \el~\cite{subramanyan2015evaluating}, invokes SAT solver~\cite{biere2013lingeling} to trim keyspace efficiently and derive the correct key. Although various logic locking approaches~\cite{roy2008epic, baumgarten2010preventing, rajendran2012security, guin2016fortis} have been proposed to obfuscate the original circuitry, SAT attack breaks all of them effectively. Furthermore, the SAT attack is also the backbone of the subsequent logic locking attacks~\cite{xu2017novel,shen2017double, limaye2021fa, zhong2022complexity}. Unlike structural attack~\cite{li2019piercing} that exploits logic redundancy to recover the secret key partially, the SAT attack relies on finding the DIP, which produces differential output for circuits with incorrect keys, analogous to test patterns that differentiate the faulty and fault-free circuits. This oracle-guided attack receives two circuits as its input, the original circuit, $C_O(X,Y)$, and its locked version, $C(X,K,Y)$. The correct key $K_c$ unlocks the circuit so that it behaves identically to the oracle, $C(X,K_c,Y)=C_O(X,Y)$, but the circuit with an incorrect key would lead to one or more output bits mismatch under certain input vectors. This discrepancy in output response, compared with the oracle, is exploited by the SAT attack. The SAT attack works in two steps, the initialization and the iterative process of pruning the key space. 

\setlength{\textfloatsep}{5pt}
\begin{algorithm}[ht]
\SetKwInOut{Input}{Input}\SetKwInOut{Output}{Output}
\Input{~Unlocked circuit, oracle ($C_O(X,Y)$) and locked circuit ($C(X,K,Y)$)}
\Output{~Correct Key ($K_c$)}
% \vspace{-5px}
\nonl \rule{0.45\textwidth}{0.4pt}

\SetAlgoLined

$i \leftarrow 1$ \;
$F\leftarrow C(X,K_{A_1},Y_{A_1})\wedge  C(X,K_{B_1},Y_{B_1})$\;
$[X_i, K_i,f] = \mathtt{sat}[F \wedge (Y_{A_i}\neq Y_{B_i})]$\;
\While{($f == \mathtt{true}$) }{
    $Y_i = \mathtt{sim\_eval}(X_i)$\;
    $F\leftarrow F\wedge C(X_i,K_{A_i},Y_i)\wedge  C(X_i,K_{B_i},Y_i)$\;
    $[X_{i+1}, K_{i+1}, f] = \mathtt{sat }[F \wedge (Y_{A_{i+1}}\neq Y_{B_{i+1}})]$\;
    $i \leftarrow i+1$ \;
}
$K_c\leftarrow K_{i}$\;
return $K_c$ \;
\caption{SAT attack on logic locking~\cite{subramanyan2015evaluating}.} \label{alg:satattack}
\end{algorithm} 

\vspace{5px}
\subsubsection{Initialization} It first constructs the miter circuit, where the locked circuit is replicated twice, $C(X,K_A,Y_A)$ and $C(X,K_B,Y_B)$, Algorithm~\ref{alg:satattack}, Line 2. The two circuits share input $X$ but not the keys $K_A$, $K_B$. Any output mismatch between the two circuits can be easily identified. In the miter circuit, the corresponding output bits from $Y_A$ and $Y_B$ are XORed and then ORed together so that a logic one at the final output indicates the output disagreement between $Y_A$ and $Y_B$ while a logic zero does not.

\vspace{5px}
\subsubsection{Pruning of key space} The attack iteratively removes the equivalence classes of incorrect keys. Since the main focus of SAT attack is the use of SAT solver to generate the appropriate input vectors, we denote the $i^{th}$ query of the SAT solver (abstracted as a function $\mathtt{sat}[\cdot]$) as the $i^{th}$ iteration of the SAT attack. At $i^{th}$ round, it finds a distinguishing input pattern $X_i$ along with assigning $f == \mathtt{true}$, where at least one output bit diverges between $C(X_i,K_{A_i},Y_{A_i})$ and $C(X_i,K_{B_i},Y_{B_i})$, $Y_{A_i}\neq Y_{B_i}$, Line 3, Line 7. The actual output $Y_i$ for this distinguishing input $X_i$ is obtained from oracle simulation, $C_O(X_i,Y_i)$, Line 5. Both $X_i$ and $Y_i$ are stored in the solver assumptions, Line 6, and carried to the subsequent iterations. {By appending this input-output pair $\{X_i,Y_i\}$ to the conjunctive normal form (CNF) in $F$, it facilitates the removal of any incorrect key combination that produces output other than the correct one $Y_i$.} The input-output pairs are accumulated so that, at the subsequent iteration, the distinguishing input pattern that the SAT solver finds not only creates differential output for the miter circuit but also satisfies all the constraint pairs $\bigwedge_{i=1,2...}\{C(X_i,K_{A_i},Y_i)\wedge  C(X_i,K_{B_i},Y_i)\}$ of the previous findings. Note that the SAT attack initializes key $K_{A_{i+1}}$ with logic values consistent with these learned IO pairs from the previous iterations. The decisions the SAT attack makes depend on the solver seeds. The SAT attack continues to eliminate the incorrect key classes and shrinks keyspace until no more distinguishing input patterns can be found, then assigns $f == \mathtt{false}$. This implies that no more incorrect keys remain. It may occur to some circuits, though rare, that more than one key is left in the key space when distinguishing input patterns no longer exists to differentiate these keys. These keys are in the equivalence class of the correct key since none would produce an output that diverges from the oracle's output. The SAT solver returns the key assignment of the last iteration as the correct key $K_c$. The detailed attack is shown in Algorithm~\ref{alg:satattack}. It is worth noting that, for every locked circuit, the very last iteration of the SAT attack always produces a UNSAT result ($f == \mathtt{false}$) where the SAT solver has exhausted all distinguishing input patterns.

%\vspace{-10px}
\section{Proposed SAT-based Test Generation Approach} \label{sec:proposed}
The test pattern returned by ATPG for detecting a \textit{sa1} (or \textit{sa0}) fault for a given node will yield one or more output differences for the faulty circuit against the fault-free one. In particular, the ATPG tool controls the faulty line with the opposite fault value and generates a test pattern where the faulty response is visible at the output. However, the ATPG tool may fail to find the appropriate input pattern during test pattern generation due to the complexity of making fault observable, like the D-Algorithm's fault activation, fault propagation, and line justification. We can broadly categorize faults as redundant and non-redundant. If the fault is redundant, no test pattern can detect it since the faulty logic does not affect the circuit's functionality. If the fault is not redundant, an input pattern must exist to propagate the fault to the output. Although ATPG may not successfully deduce a test pattern, it does not necessarily say that the fault is redundant. The fault could still belong to either group, redundant or non-redundant. The focus of this section is to generate test patterns for non-redundant faults and, at the same time, separate the redundant faults. \textit{In particular, this section presents how to precisely label a fault as redundant or not when ATPG fails to give the test pattern for an undetected fault or determines the appropriate test pattern in concurrence with the identification of a hard-to-detect fault.} We introduce a novel approach to construct an equivalence mapping between test pattern generation of stuck-at faults and the SAT attack on logic locking. Our fault modeling inserts key gates at the faulty lines so that both fault observability and controllability are fulfilled when the SAT attack tries to find distinguishing patterns to decrypt the key bits. For redundant faults, our model returns UNSAT at the first iteration of the SAT attack without any distinguishing input pattern, indicating that no pattern could make the fault observable. For any hard-to-detect fault, the SAT attack obtains a satisfiable input assignment at the first round, which is the desired test vector. Our approach offers a solution for the test pattern generation problem of hard-to-detect faults, in a novel perspective from logic locking, where any patterns derived from the SAT attack are the ones we needed in the test pattern generation domain.

\vspace{-10px}
\subsection{Novel miter construction for stuck-at fault with key-dependent circuit in logic locking}\label{sec:modeling}
The SAT attack on logic locking has shown tremendous success in deriving the correct key of various locking in a few seconds~\cite{subramanyan2015evaluating}. As described in Section~\ref{sec:satattack}, this means that the SAT attack found the input patterns necessary for removing all incorrect key combinations within the recorded time frame. For example, the locked benchmark of \textit{c880} with 192-bit key from random logic locking (randomly inserts XOR/XNOR key gates) is broken by the SAT attack using only 30 distinguishing input patterns in less than 1 second. The efficiency of SAT attack, in terms of both attack time and the number of input patterns, motivates us to exploit it to identify any non-redundant faults and generate the associated test patterns for these hard-to-detect faults that are not previously detected by a commercial ATPG tool. Moreover, it is also desired if we can simultaneously determine any undetected faults that are redundant. To equivalently transform a circuit with stuck-at faults to a locked one, we need to make sure that the properties of these faults are controllable and observable when the SAT attack derives distinguishing input patterns. This section presents the applying of novel miter construction in the SAT attack for test generation of stuck-at faults.

%  \begin{figure}[t]
%     \centering 
%     \includegraphics[width=0.8\columnwidth]{LLgate.pdf}
%     \caption{Challenge in modeling an \textit{saf} using XOR gates. (a) an interconnect with a stuck-at fault (b) XOR gate with input logic 0 (c) XOR gate with input logic 1. } \label{fig:xorgate} 
%     % \vspace{-10px}
% \end{figure}

As logic locking uses XOR key gates, the question that first comes to mind is whether it is possible to model a stuck-at fault using an XOR gate. Unfortunately, we cannot model a stuck-at fault due to the symmetric nature of the XOR gate. If we model a stuck-at fault with an XOR gate as a key, either a logic 0 or 1 at the input can propagate the key to its output. However, during ATPG of a stuck-at 0 \textit{sa0}, or stuck-at 1 (\textit{sa1}), a logic 1 or 0, should be placed on the fault site to activate it. This made it impossible for XOR-based locking to model \textit{saf} as both patterns are valid DIPs for the key bit.

\begin{figure}[t]
    \centering  
    \includegraphics[width=.85\columnwidth]{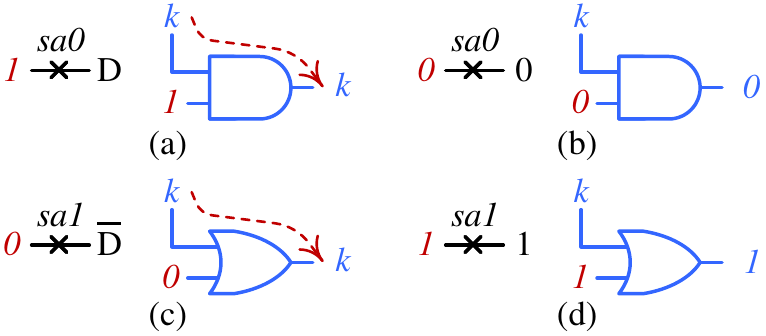}\vspace{-5px}
    \caption{Logic locking-based modeling of a \textit{saf} with AND or OR key gate. Converting a \textit{sa0} to AND key gate, (a) successful propagation of key $k$ with logic 1, (b) failed propagation of $k$ with logic 0; a \textit{sa1} to OR key gate, (c) successful propagation of $k$ with logic 0, and (d) failed propagation of $k$ with logic~1.} \label{fig:sa0and} 
\end{figure}

As XOR/XNOR key gates can not be applied inside the miter construction in generating a test for a stuck-at fault, \textit{we need to find a different key gate so that its input can only have the opposite value of the stuck-at fault for key propagation.} Logic 1 complements \textit{sa0} fault for making it observable, while logic 0 does not change the functionality. If we pick AND gate, instead of XOR, as the key gate, logic 1 at the input of the key gate will help $k$ to the key gate's output, as shown in Figure~\ref{fig:sa0and}(a), but a logic 0 at the input blocks the key propagation with a constant 0 at the output, as shown in Figure~\ref{fig:sa0and}(b). This means that, if a DIP exists, the SAT attack will assign logic 1 to the input of the key gate for the miter circuit since logic 0 could not fulfill the differential output condition between the incorrect and correct keys. The input vector used for key derivation in the SAT attack satisfies the controllability and observability requirement of \textit{sa0} in test pattern generation. 

For \textit{sa1} fault, we need to assign logic 0 at the fault site for observing the \textit{sa1} since having the same logic as the fault, logic 1, impedes it from being revealed. Analogous to selecting AND key gate for \textit{sa0} test pattern generation, we choose OR gate to represent \textit{sa1}. A logic 0 at the input of the OR key gate allows $k$ to appear at the key gate's output, as shown in Figure~\ref{fig:sa0and}(c). However, placing a logic 1 at the input blocks the key visibility with a constant 1 at the output, as shown in Figure~\ref{fig:sa0and}(d). Similar to the analysis on AND key gate and \textit{sa0}, for generating DIPs, the SAT attack must assign logic 0 to the input of the key gate because logic 1 fails to differentiate the circuit's output between the incorrect and the correct key bit.

 \begin{figure}[ht]
    \centering 
    \includegraphics[width=\columnwidth]{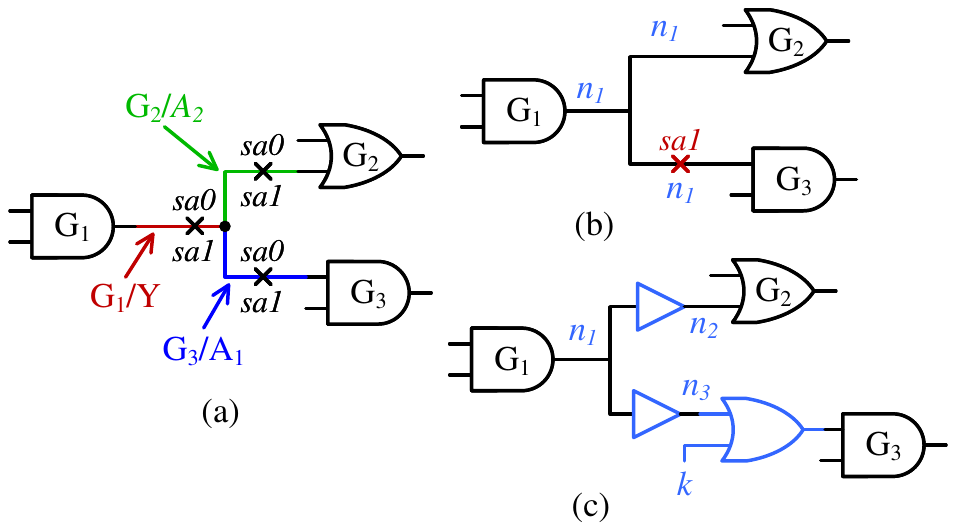} \vspace{-15px}
    \caption{Stuck-at-faults in the presence of multiple fanout branches. (a) Fanout branch naming with TetraMAX. (b) Naming convention used in \textit{.bench} file. (c) Fanout branch renaming using buffers.} \label{fig:buffer} 
    %  \vspace{-10px}
\end{figure} 

To address the stuck-at fault detection at fanouts, we need to rename the fanout segment in the \textit{.bench} file used by the SAT attack. Note that any synthesized netlist from a commercial tool considers each fanout segment with a unique name, which is tied to either the input or the output of a gate. Figure~\ref{fig:buffer}(a) shows the fanout where the output of gate $G_1$ is connected to both the inputs of $G_2$ and $G_3$. TetraMAX inserts faults, both \textit{sa1} and \textit{sa0}, at all the fanout segments, named as $G_1/Y, G_2/A_2$, and $G_3/A_1$, respectively. However, as these three segments share the same logic value, in the \textit{.bench} file, all these segments will be treated as a single node, say $n_1$. As a result, we cannot add faults to the green or blue segments only. However, the SAT attack requires the bench file as input, all the fanout branches have the same name, as shown in Figure~\ref{fig:buffer}(b). So, we can only add two faults instead of six. To address this problem, we added one buffer to rename the fanout branches. Figure~\ref{fig:buffer}(c) shows the equivalent locked circuit of a \textit{sa1} fault at the first input of gate $G_3$.

Based on the above analysis, any \textit{sa0} or \textit{sa1} can be converted to its AND key gate or OR key gate equivalent in logic locking while preserving the observability of stuck-at fault. Note that, aside from generating DIPs, the SAT attack also derives the key value. From the logic locking perspective, the correct key decrypts the locked circuit so that it is functionally the same as the oracle, $C(X,K_c,Y)=C_O(X,Y)$. For the AND key gate, the logical value on the wire (before locking) can pass through the key gate unmodified with key $k=1$ but assigning key $k=0$ forces the AND output to constant 0, which alters the original circuit functionality. Likewise, with OR as the key gate, the correct key is $k=0$, while the output of the key gate will be kept at constant 1 for the incorrect key value $k=1$. Hence, in addition to DIPs, our equivalent representation of stuck-at faults can be further confirmed by checking the correct key $k=1$ for all \textit{sa0} and $k=0$ for \textit{sa1} faults.

\vspace{-10px}
\subsection{Identification of Redundant Faults} \label{sec:redundant}
Our proposed miter construction with the SAT attack for test generation can also identify any redundant faults. If a stuck-at fault (either \textit{sa0} or \textit{sa1}) is redundant, no pattern can ever propagate this fault since it is not influencing the circuit's functionality. The output behaves the same for the faulty and fault-free circuits. In the same way, when we turn the redundant fault to its equivalent locked circuit, the key cannot be observed from the output as well, as it is located at the redundant line without affecting the primary output. % \todoYadi{Talk about satisfiability. The circuit will be satisfiable for both the key values (k=0 and k=1). You can drwa a miter circuit indicating only one key, and update the subsequent discussion.} 
 \begin{figure}[!ht]
    \centering  \vspace{-10px}
    \includegraphics[width=0.8\columnwidth]{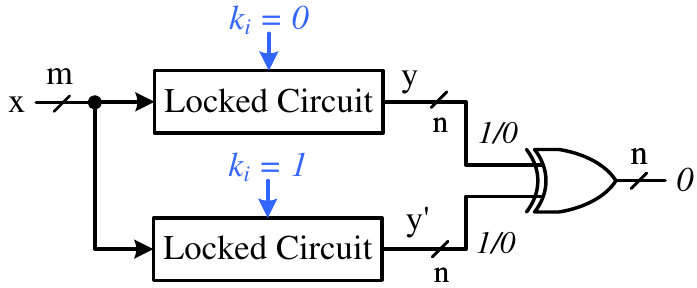}%\vspace{-5px}
    \caption{SAT attack's miter circuit with 1-bit key $k_i$ for $i^{th}$ \textit{saf} (redundant fault).} \label{fig:miterckt} 
     \vspace{-5px}
\end{figure}

Let us assume that the $i^{th}$ stuck-at fault in the circuit is redundant. Once we lock it with the appropriate key gate (and buffer if needed) with 1-bit key $k_i$, we invoke the SAT attack in an attempt to find a DIP for this fault. Since the SAT attack constructs the miter circuit to search for the DIP, as illustrated in Figure~\ref{fig:miterckt}, both locked circuits share the common \textit{m}-bit input $x$. The \textit{n}-bit outputs, $y$ and $y'$, are XORed in a bit-by-bit manner. As the $i^{th}$ fault is redundant, the key bit $k_i$ has no impact on the n-bit output $y$, and the two locked circuits have an identical response for $k_i=0$ and $k_i=1$. For test pattern generation, the miter circuit would produce the exact same output $y=y'$ under any input combinations. As a result, the miter output is always zero, and no output difference between $y$ and $y'$ can be observed. This means that the SAT attack could not find any DIP to differentiate $k_i=0$ and $k_i=1$, and it would reach the UNSAT conclusion at the first query of SAT solver on Line 3, Algorithm~\ref{alg:satattack}. Note that, \textit{as redundant faults do not change the circuit's functionality, we can ignore them during fault coverage computation if identified correctly.}

\vspace{-10px}
\subsection{ATPG using the SAT Attack on Logic Locking} \label{sec:testpatterngen} %SAT-based Logic Locking Attack to find the key

Just as a few test patterns from the ATPG tool could expose multiple stuck-at faults, the SAT attack can also rule out the exponential number of incorrect keys with a few distinguishing patterns. When it comes to test pattern generation for hard-to-detect faults, we have the option to select how many of these faults we can analyze together. The conservative approach is to generate a test pattern for every fault. This approach can also identify whether a fault is redundant or not by checking if the SAT attack returns a DIP. On the other hand, we can combine the equivalent conversion of multiple faults in one locked circuit with the same number of key gates as the faults. We then ask the SAT attack to break this locked circuit and collect all the DIPs. As the SAT attack generally trims multiple incorrect keys from the search space with only a few DIPs, analyzing a group of faults has the potential of reduced pattern set than inspecting one fault at a time. Both strategies work for any stuck-at fault, regardless of being redundant or not. In the following sections, we present a comprehensive discussion of both approaches by focusing on undetected faults. The first approach asks the SAT attack for a DIP on every undetected fault, while the second one targets a group of faults so that the SAT attack solves key bits simultaneously.

\vspace{5px}
\subsubsection{Approach 1 -- Generate One Test Pattern per Fault} \label{sec:onetestperfault}

Approach 1 focuses on finding a single test pattern for an undetected stuck-at fault using the SAT attack. The equivalent locked circuit contains a 1-bit key. To solve the 1-bit key, the SAT attack only needs to query the SAT solver twice. At the first query, Line 3 of Algorithm~\ref{alg:satattack}, the SAT solver returns the input pattern where the primary output differs for the correct and incorrect key assignments, that is, between logic 0 and logic 1. This input pattern, along with the corresponding output, simulated from the oracle, is saved in the IO constraints $F$. Note that the wrong key bit is implicitly removed from the search space as it does not satisfy the IO pair stored in $F$. Only the correct key bit matches the IO behavior in $F$, and it is the only candidate that remains in the search space. With constraint $F$ appended in the satisfiability of the miter circuit (Algorithm~\ref{alg:satattack}, {Line 6}), the SAT solver must return UNSAT at the second query, and it could not produce any differential output when no more incorrect keys exist in the key space. The second scenario is that the SAT attack could not find any distinguishing input pattern to differentiate the keys in the search space at the first query of SAT solver, and it terminates the while loop (Algorithm~\ref{alg:satattack}, Lines 4-9). It also returns the hypothesis key $K_i$, but it may not align with the correct key value discussed in the novel miter construction for stuck-at faults, Section~\ref{sec:modeling}. This is caused by the fault at the redundant line where faulty value can never reach the output ports, and no input pattern can be found. By including one fault at a time, we can quickly identify which fault is redundant by determining whether DIP is obtained from the SAT attack.  

%to adjust spacing
 \begin{figure}[!ht]
    \centering 
    \includegraphics[width=1\linewidth]{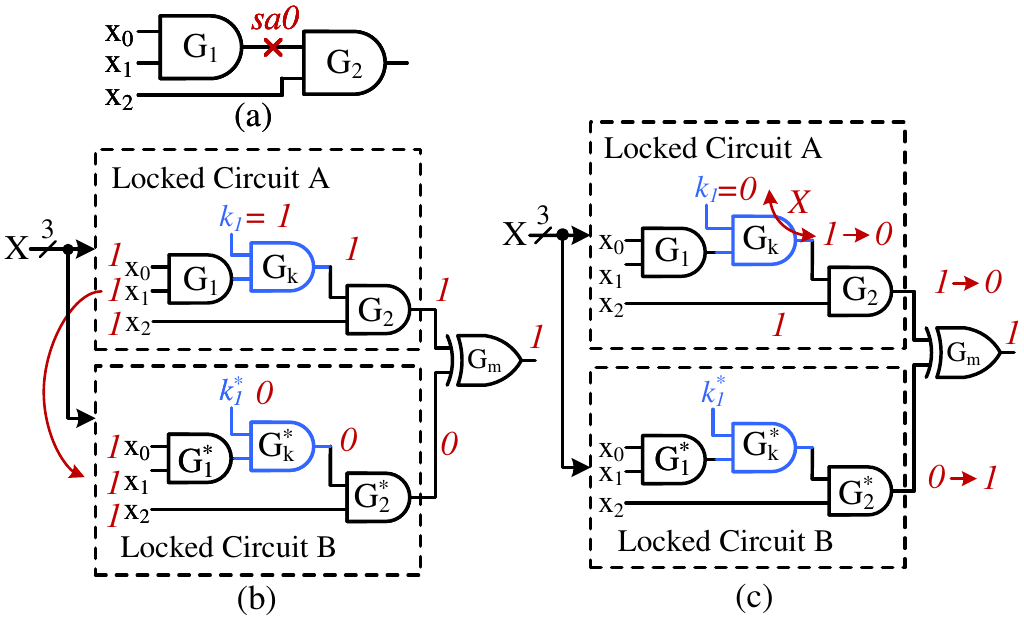} \vspace{-15px}
    \caption{The proposed test pattern generation with the SAT attack miter. (a) A simple circuit with a \textit{sa0} fault. (b) No backtrack in deriving the satisfiable assignment with $k_1=1$. (c) Backtrack at an earlier stage with $k_1=0$. 
    } \label{fig:approach1}  %\vspace{-10px}
\end{figure}

Compared to \cite{fujita2014efficient}, our proposed approach can determine the test patterns without conflicts or resolve conflicts at an earlier stage due to the initialization of keys in one locked circuit inside the miter. Figure~\ref{fig:approach1}(a) shows the example circuit with a \textit{sa0} fault. Note that two possible scenarios exist for our proposed approach where the SAT attack assigns $k_1$ to logic 1 or 0 at the start. Let us first consider when the SAT attack assigns $k_1=1$, denotes as \textit{Case 1}) and is shown in Figure~\ref{fig:approach1}(b). Applying the same assumption mentioned in Section~\ref{subsec:background-ATPG}, the SAT solver assigns the first input of $G_m$ to logic 1. All the literals in the miter can be iteratively implied without raising conflict or backtracking for CNF clauses, and a test is found. If the SAT attack starts with $k_1=0$, denoted as \textit{Case 2} and shown in Figure~\ref{fig:approach1}(c), a conflict arises at the output of key gate $G_k$ and is resolved locally by conflict-driven clause learning (CDCL)~\cite{biere2021handbook} without having to trace back the entire miter circuit like~\cite{fujita2014efficient}. As a result, the output of circuit A is reassigned to 0 at an earlier stage~\cite{fujita2014efficient}. This ensures the SAT solver backtrack at a much earlier stage to determine the satisfiability, and a hard-to-detect fault can be found efficiently.

 \begin{figure}[t]
    \centering 
    \includegraphics[width=1\linewidth]{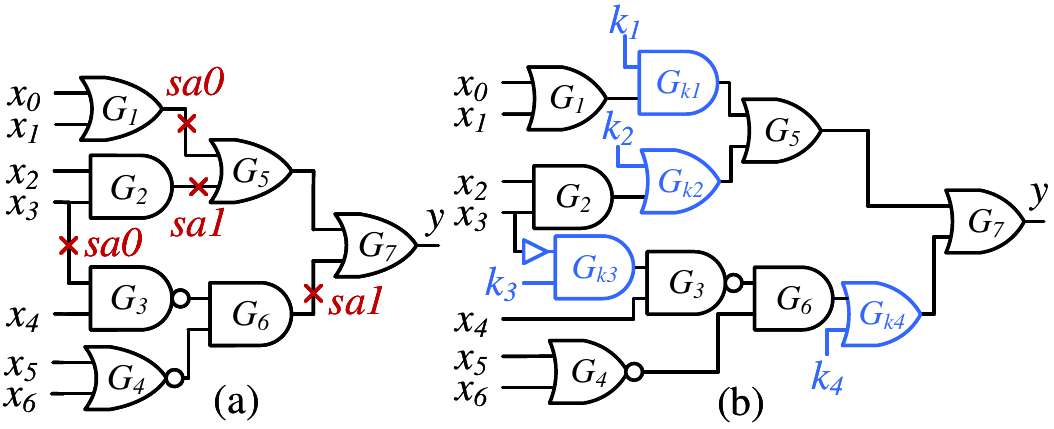} \vspace{-10px}
    \caption{Test generation with a group of faults. (a) circuit with multiple faults (b) the \textit{saf}s equivalence with logic locking. } \label{fig:approach2} %\vspace{-10px}
\end{figure}

% \vspace{5px}
\subsubsection{Approach 2 -- Generate Test Patterns for a Group of Faults} \label{subsubsec:approach-2}

While the first approach details the test pattern generation considering a single fault, this Approach 2 targets multiple faults simultaneously. Instead of adding one key gate per locked circuit as in Approach 1, the second approach locks a circuit with multiple key gates where the number of keys is the same as the to-be-analyzed faults. This approach resembles the prevalent strategy within the logic locking community, where a circuit is locked with multiple key bits. From the SAT attack perspective, each distinguishing pattern can, in general, remove multiple incorrect keys in the search space. This is because any incorrect key assignment in the traditional logic locking techniques is more likely to produce the wrong and corrupted output on a given input vector. This statement also applies to our proposed logic locking conversion of stuck-at faults with AND/OR key gates at the fault sites (as illustrated in the example below). In addition, our key-dependent fault-equivalence conversion supports the test pattern generation of detecting both \textit{sa0} and \textit{sa1} of the same fault site, where one can simply insert two key gates, an AND and OR gate each, in serial at the target locations. The goal of this approach is to reduce the number of test patterns. Following the novel miter construction for faults in Section~\ref{sec:modeling}, we transform $n$ hard-to-detect faults into its equivalent locked circuit with $n$ key bits. We collect all the DIPs the SAT solver identifies and the key value from the SAT attack. When faults are detectable, one or more input patterns always exist to differentiate the correct key from the wrong one. It may be true that some faults could be situated at redundant lines, where no test pattern can be generated since the circuit output does not depend on these faults. To know precisely how many faults are detected through the SAT attack's distinguishing pattern, we run the fault simulation with the target group of faults and the extracted patterns. 

\begin{table}[ht]
\caption{The SAT attack uses only 2 patterns to eliminate the incorrect keys from the search space. If the output matches the correct output, we put \cmark, else \xmark.}\vspace{-10px}
\begin{center}
\begin{tabular}{ |M{2.5cm}|M{2.5cm}|M{2.5cm}| } 
 \hline
 4-bit key & Pattern 1 & Pattern 2\\
 $\{k_1,...,k_4\}$ & $\{1001100\}$ & $\{0001100\}$\\ \hline
 0000 & \cmark & \xmark \\ \hline
 0001 & \cmark & \xmark \\ \hline
 0010 & \xmark & \cmark \\ \hline
 0011 & \cmark & \xmark \\ \hline
 0100 & \cmark & \xmark \\ \hline
 0101 & \cmark & \xmark \\ \hline
 0110 & \cmark & \xmark \\ \hline
 0111 & \cmark & \xmark \\ \hline
 1000 & \cmark & \xmark \\ \hline
 1001 & \cmark & \xmark \\ \hline
 {\color{red}{\textbf{1010}}} & {\color{red}\cmark} & {\color{red}\cmark} \\ \hline
 1011 & \cmark & \xmark \\ \hline
 1100 & \cmark & \xmark \\ \hline
 1101 & \cmark & \xmark \\ \hline
 1110 & \cmark & \xmark \\ \hline
 1111 & \cmark & \xmark \\  \hline
\end{tabular}
\end{center}
\label{tab:truthtable}\vspace{-10px}
\end{table} 

We take the combinational circuit with 4 stuck-at faults, shown in Figure~\ref{fig:approach2}(a), as an example. Both \textit{sa0}s are turned to AND gates $G_{k1},G_{k3}$ with keys $k_1,k_3$, and the \textit{sa1}s are locked with OR gates $G_{k2},G_{k4}$ with keys $k_2,k_4$. Note that the input $x_3$ branches to two lines and only the wire connected to the input of gate $G_3$ has \textit{sa0}, but not for the one at the input of gate $G_2$. So, we include one buffer for the conversion of this \textit{sa0}, as described in Section~\ref{sec:modeling}. With Approach 2, the SAT attack only uses 2 DIPs, $P_1 = \{x_0,x_1,...x_6\}=\{1001100\}$, $P_2 = \{x_0,x_1,...x_6\}=\{0001100\}$, instead of 4 DIPs with Approach 1, to prune all 15 wrong key combinations and correctly derive the key vector $\{k_1,k_2,k_3,k_4\}=\{1010\}$. For both patterns, we show in Table~\ref{tab:truthtable} detailing whether the 16 possible keys produce the correct output $y$ or not, where \cmark~indicates a match with the oracle output and \xmark~for a mismatch, and the correct key is highlighted in red. On the first iteration of the SAT attack, %\todoYadi{make sure that you pass the texts through grammar check} 
it finds the first distinguishing pattern $P_1 =\{1001100\}$, which removes key $\{k_1,...,k_4\}=\{0010\}$ from the key space. On the second iteration, it returns another distinguishing pattern $P_2 =\{0001100\}$ that crosses out another 14 keys, leaving only one key in the key space. On the third iteration, no more DIP can be found to create differential output with the only remaining key $\{k_1,...,k_4\}=\{1010\}$ as the SAT attack already removed all the incorrect keys at the first two iterations. As it does not finds any satisfiable pattern in the third iteration, the SAT attack terminates and returns the correct key $\{k_1,...,k_4\}=\{1010\}$.

The proposed Approach 2 can significantly reduce the test pattern generation time compared to Approach 1. As the miter construction in the SAT attack initialize the key inside the locked circuit (e.g., instance A, see Figure~\ref{fig:approach1}) and is assigned with Boolean logic consistent with the learned clauses of the previous rounds, it offers $n$ locations potentially for reducing the conflicts or backtracks for the SAT solver when $n$ faults are grouped together. This can offer much more efficiency to deduce the satisfiable assignment as the miter circuit of the single fault has 1 initial starting point while the $n$-bit key provides $n-1$ more pre-assigned locations. This leads to further test time reduction compared with a single stuck-at fault per run. This observation is verified in Table~\ref{tab:test-pattern-generation-time} in Section IV by comparing the ratio of Approach 2 to Approach 1 on the test time of hard-to-detect fault list ${DF_P}$. The total time can be much smaller than the time required to generate a single pattern in Approach 1 for $b\_19$ benchmark circuit. An increased test time gain can be obtained when more faults are transformed together in one locked circuit.

Approach 2 can also identify redundant faults. If all the $n$-faults are redundant, the SAT-attack tool will not return any DIPs and will identify them uniquely. However, if there exist one or more detectable faults, we cannot identify them uniquely from the UNSAT conclusion. One more step is necessary to identify the redundant faults using fault simulation. The patterns obtained from the proposed Approach 2 need to be applied to the fault simulator and identify the undetected faults. As no test pattern exists for any redundant faults, all these undetected faults reported by the fault simulator must be redundant. In summary, Approach 2 can target a group of redundant faults where the SAT attack can finish solving all keys in the first few iterations with the UNSAT conclusion at the very last round. 

% changed Approaches 1 and 2 as functions
\setlength{\textfloatsep}{5pt}
\begin{algorithm}[t]
\SetKwInOut{Input}{Input}\SetKwInOut{Output}{Output}
\SetKwProg{ApOne}{$\mathtt{Approach}\text{--}\mathtt{1}$}{ is}{end}
\SetKwProg{ApTwo}{$\mathtt{Approach}\text{--}\mathtt{2}$}{ is}{end}
\Input{~Combinational circuit in \textit{.bench} format ($C_N$) and standard cell library ($stdlib$)}
\Output{~Redundant fault set ($RF$), hard-to-detect fault set ($DF$), test pattern set ($P_{A1}$ and $P_{A2}$)}

\nonl \rule{0.45\textwidth}{0.4pt}

\SetAlgoLined
 % use pseudo-code
% \vspace{5px}
\nonl //-------- Step-1: Generate undetected fault list $L$ --------\\
% \vspace{5px}
$C_{IN} \leftarrow \mathtt{fromBench}(C_N)$ \;
$C_{DN}\leftarrow\mathtt{toTechDependentNetlist}(C_{IN},stdlib)$ \;

$tp\leftarrow \mathtt{write\_drc\_file}(C_{DN},stdlib)$ \;
$\mathtt{loadATPG}(C_{DN},stdlib,tp)$ \;
$\mathtt{addFaults}(\textit{sa0},\textit{sa1},\textit{`all'})$ \;
$\mathtt{setAbortLimit}(\textit{max})$ \;
$\mathtt{runATPG}$ \;
$L \leftarrow \mathtt{reportFault}$(\textit{undetected}) \;%Save all untestable faults to a list $L$ \;

$[RF_{A1}, DF_{A1}, P_{A1}] \leftarrow \mathtt{Approach}\textbf{--}\mathtt{1}(C_{DN}, C_N, L)$ \;

$[RF_{A2}, DF_{A2}, P_{A2}] \leftarrow \mathtt{Approach}\textbf{--}\mathtt{2}(C_{DN}, C_N, L)$ \;
return $RF_{A1,A2}$, $DF_{A1,A2}$, $P_{A1,A2}$ \;

\vspace{5px}
\nonl //-------- Step-2: Approach 1 --------------------------------\\
% \vspace{5px}
\textbf{function} \ApOne{($C_{DN}, C_N, L$)}{

 $[RF_{A1},DF_{A1},P_{A1}] \leftarrow \varnothing$ \;
\For{$i \leftarrow 1\ \KwTo\  |L|$}{
    $f \leftarrow L[i]$ \; 
    $C_{L_{A1}}\leftarrow \mathtt{locked}\textbf{--}\mathtt{ckt}(C_{DN},f,\textit{`Approach 1'})$ \;
    $C_{B_{A1}}\leftarrow\mathtt{toBench}(C_{L_{A1}})$ \;
    $[p,k]\leftarrow \mathtt{SAT}\text{-}\mathtt{attack}(C_{B_{A1}},C_N)$ \;
    \uIf{$p=\varnothing$}{
        $RF\leftarrow \mathtt{append}(f)$ \;
    }\ElseIf{$p!=\varnothing\ \&\ k=k_{ref}$}{
        $DF\leftarrow \mathtt{append}(f)$ \;
        $P_{A1} \leftarrow \mathtt{append}(p)$
    }
}
return $RF_{A1}$, $DF_{A1}$, and $P_{A1}$ \;
}
\vspace{5px}
\nonl //-------- Step-3: Approach 2 --------------------------------\\
% \vspace{5px}
\textbf{function} \ApTwo{($C_{DN}, C_N, L$)}{

 $[RF_{A2},DF_{A2},P_{A2}] \leftarrow \varnothing$ \;

    $C_{L_{A2}}\leftarrow \mathtt{locked}\textbf{--}\mathtt{ckt}(C_{DN},L,\textit{`Approach 2'})$\;

    $C_{B_{A2}}\leftarrow\mathtt{toBench}(C_{L_{A2}})$ \;
        
	$[P,K]\leftarrow \mathtt{SAT}\text{-}\mathtt{attack}(C_{B_{A2}},C_N)$ \;
    $[DF_{A2}, RF_{A2}] \leftarrow \mathtt{faultSim}(C_{DN},L,P)$ \;
    $P_{A2} \leftarrow P$ \;
% \vspace{5px}
return $RF_{A2}$, $DF_{A2}$, and $P_{A2}$ \;
}
\caption{Proposed SAT-based test pattern generation for hard-to-detect faults and identification of redundant faults.} \label{alg:TP-generation}
\end{algorithm}

%to adjust placing
\begin{table*}[t]
\caption{Stuck-at Faults Summary for ITC'99 Benchmarks.}\vspace{-10px}
\small
\begin{center} %\setlength{\tabcolsep}{0.55em}
\begin{tabular}{|c|c|c|c|c|c|c|c|c|c|c|c|c|}
\hline
\multirow{2}{*}{\textbf{Benchmark}}   &    \multirow{2}{0.8cm}{\centering \textbf{Gate Count}}  & \multirow{2}{1.7cm}{\centering \textbf{ Total Faults ($\bm{TF}$)}} & \multirow{2}{0.8cm}{$\bm{DF_{\textit{TMAX}}}$} & \multicolumn{6}{c|}{\textbf{TetraMAX II}} & \multicolumn{2}{c|}{\textbf{PA}} & \textbf{Total}\\ \cline{5-12}
&& &     & $\bm{PT}$ & $\bm{UD}$   & $\bm{AU}$  & $\bm{ND}$   & $\bm{UF}$ & $\bm{FC}$ \textbf{(\%)} & $\bm{RF_P}$   & $\bm{DF_P}$ & $\bm{FC_T}$\hspace{-3px} \textbf{(\%)}\\ \hline
b04\_opt\_C & 543    & 3554    & 3549    & 0  & 5    & 0   & 0    & 5    & 99.86 & 5    & 0  & 100 \\ \hline
b04\_C      & 657    & 4144    & 4094    & 0  & 50   & 0   & 0    & 50   & 98.79 & 50   & 0  & 100 \\ \hline
b05\_opt\_C & 505    & 3272    & 3265    & 4  & 3    & 0   & 0    & 7    & 99.85 & 3    & 4  & 100 \\ \hline
b05\_C      & 943    & 5850    & 4747    & 0  & 1099 & 4   & 0    & 1103 & 81.15 & 1103 & 0  & 100 \\ \hline
b07\_opt\_C & 371    & 2456    & 2455    & 0  & 1    & 0   & 0    & 1    & 99.96 & 1    & 0  & 100 \\ \hline
b07\_C      & 385    & 2470    & 2464    & 0  & 6    & 0   & 0    & 6    & 99.76 & 6    & 0  & 100 \\ \hline
b11\_opt\_C & 511    & 3318    & 3316    & 0  & 2    & 0   & 0    & 2    & 99.94 & 2    & 0  & 100 \\ \hline
b11\_C      & 734    & 4378    & 4212    & 0  & 161  & 0   & 5    & 166  & 96.21 & 161  & 5  & 100 \\ \hline
b12\_opt\_C & 886    & 6048    & 6047    & 0  & 1    & 0   & 0    & 1    & 99.98 & 1    & 0  & 100 \\ \hline
b13\_C      & 290    & 1928    & 1848    & 0  & 80   & 0   & 0    & 80   & 95.85 & 80   & 0  & 100 \\ \hline
b14\_opt\_C & 9811   & 58584   & 58265   & 0  & 318  & 0   & 1    & 319  & 99.46 & 319  & 0  & 100 \\ \hline
b14\_C      & 5477   & 35844   & 35806   & 0  & 33   & 2   & 3    & 38   & 99.89 & 38   & 0  & 100 \\ \hline
b15\_opt\_C & 7206   & 48220   & 46922   & 19 & 1054 & 38  & 187  & 1298 & 97.33 & 1287 & 11 & 100 \\ \hline
b15\_C      & 8462   & 53470   & 51952   & 0  & 1329 & 76  & 113  & 1518 & 97.16 & 1518 & 0  & 100 \\ \hline
b17\_opt\_C & 23523  & 157418  & 154248  & 15 & 1153 & 300 & 1702 & 3170 & 97.99 & 3144 & 26 & 100 \\ \hline
b17\_C      & 31091  & 192174  & 187897  & 0  & 3866 & 33  & 378  & 4277 & 97.77 & 4276 & 1  & 100 \\ \hline
b20\_opt\_C & 12170  & 79748   & 79644   & 0  & 99   & 0   & 5    & 104  & 99.87 & 104  & 0  & 100 \\ \hline
b20\_C      & 19792  & 118298  & 117614  & 35 & 631  & 18  & 0    & 684  & 99.44 & 631  & 53 & 100 \\ \hline
b21\_opt\_C & 12344  & 80504   & 80398   & 0  & 95   & 0   & 11   & 106  & 99.87 & 106  & 0  & 100 \\ \hline
b21\_C      & 20109  & 120436  & 119742  & 8  & 684  & 0   & 2    & 694  & 99.43 & 686  & 8  & 100 \\ \hline
b22\_opt\_C & 17614  & 114556  & 114387  & 1  & 165  & 0   & 3    & 169  & 99.85 & 169  & 0  & 100 \\ \hline
b22\_C      & 29316  & 175510  & 174653  & 33 & 795  & 29  & 0    & 857  & 99.52 & 796  & 61 & 100 \\ \hline
b18\_opt\_C & 71392  & 469602  & 469044  & 0  & 479  & 17  & 62   & 558  & 99.88 & 544  & 14 & 100 \\ \hline
b18\_C      & 112421 & 672242  & 668478  & 0  & 3658 & 96  & 10   & 3764 & 99.44 & 3760 & 4  & 100 \\ \hline
b19\_C      & 226936 & 1355584 & 1347025 & 6  & 8236 & 191 & 126  & 8559 & 99.37 & 8460 & 99 & 100 \\ \hline
\end{tabular}
 \end{center}\label{tab:itc} \vspace{-10px}
\end{table*}

\vspace{-10px}
\subsection{Test Pattern Generation and Redundant Fault Identification Algorithm}\label{sec:algorithm}
Algorithm~\ref{alg:TP-generation}  shows the identification of redundant faults and the test pattern generation process for hard-to-detect faults. The algorithm has three steps, and it first performs traditional ATPG using a commercial tool to generate test patterns and report undetected faults. As our objective is to use the SAT attack to generate test patterns and identify redundant faults, we made a few adjustments to the traditional approach of test pattern generation, which starts from synthesizing a circuit using a commercial tool (e.g., Synopsys Design Compiler). 
If we are given an RTL code, we follow the traditional approach to obtain the technology-dependent gate-level netlist from design synthesis with standard cell library \textit{stdlib}. If the synthesized netlist is a sequential design, scan-chain insertion is required to convert the sequential design to a combinational one so that ATPG can generate test patterns efficiently. On the other hand, if the benchmark is already in the combinational \textit{bench} format, \eg the \textit{`\_C'} circuits in the ITC'99 benchmark suite (\url{https://github.com/squillero/itc99-poli}), we can directly convert the \textit{bench} file $C_{N}$ to a technology-independent gate-level netlist $C_{IN}$, Algorithm~\ref{alg:TP-generation}, Line 1. Then, this technology-independent netlist $C_{IN}$ can be mapped to a technology-dependent netlist $C_{DN}$ with standard cell library $stdlib$, Line 2. This can be done without synthesizing the design $C_{IN}$, which may introduce potential line mismatch during optimization, and the synthesized netlist may deviate from its original \textit{bench} netlist $C_N$. Any standard cell library can map the technology-independent netlist to a technology-dependent one for commercial ATPG tools. As we target only the stuck-at faults, they are independent of the parameters in the library, unlike delay, bridging faults, or resistive opens that are dependent on the intrinsic properties of the technology node. The ATPG tool also requires a test protocol $tp$ in \textit{SPF} format, which can be either generated from netlist synthesis or directly written within the ATPG tool~\cite{SynopsysTetraMAX} by the command $\mathtt{write\_drc\_file}$, Line 3. After loading netlist $C_{DN}$, library $stdlib$ and test protocol $tp$ to ATPG tool, Line 4, stuck-at 0 (\textit{sa0}) and stuck-at 1 (\textit{sa1}) faults are assigned to all lines in the circuit, including the primary input and output, Line 5. Since fault coverage can be improved by increasing the allotted number of backtracks and remade decisions of the ATPG tool, we set the abort limit to its maximum value, Line 6. ATPG is then invoked to run test pattern generation and fault coverage analysis, Line 7, and report any undetected faults by the tool to a list $L$, Line 8.

\vspace{-10px}
The algorithm identifies the redundant faults from the undetected fault list $L$ (Lines 9, 12-27) in Step-2 using Approach 1. Three empty sets are initialized, hard-to-detect fault set $DF_{A1}$, redundant fault set $RF_{A1}$, and test pattern set $P_{A1}$, Line 13. For each fault $f$ in the undetected list $L$ (Line 15), the locked circuit $C_{L_{A1}}$ is modeled with a single key bit, Line 16. After converting the locked netlist $C_{L_{A1}}$ to \textit{bench} format $C_{B_{A1}}$, Line 17, the SAT attack is executed with $C_{B_{A1}}$ and the oracle $C_N$ to obtain the DIP $p$ and the key value $k$, Line 18. If the SAT attack does not return a DIP from the miter circuit, $p$ is empty, and the fault is redundant, where it is added to the redundant list $RF_{A1}$, Lines 19-20. However, if the SAT attack finds a DIP as well as the correct key value compared to the reference key ($k_{ref}=1$ for \textit{sa0} and $k_{ref}=0$ for \textit{sa1}) for the proposed fault modeling, Line 21, this fault $f$ is detected. It is appended to the hard-to-detect list $DF_{A1}$ and its DIP $p$ is added to test pattern set $P_{A1}$, Lines 22-23. Note that fault $f$ belongs to either category, $RF_{A1}$ or $DF_{A1}$, and no fault skips the \textit{if-else-if} statement, as analyzed in Section~\ref{sec:modeling},~\ref{sec:redundant}.

In Step-3, the algorithm optimizes the test pattern set for all undetected faults in list $L$, Lines 10, 28-36, as the previous step reports either one test pattern or none per fault. The hard-to-detect fault set $DF_{A2}$, redundant fault set $RF_{A2}$, test pattern set $P_{A2}$ is initialized as an empty set, Line 29. All faults in the undetected list $L$ are converted to a locked circuit $C_{L_{A2}}$ with $|L|$ number of key gates, Line 30, as described in Section~\ref{subsubsec:approach-2}. The locked circuit $C_{L_{A2}}$ is then mapped it to its equivalent \textit{bench} file $C_{B_{A2}}$, Line 31. Both $C_{B_{A2}}$ and $C_N$ are applied to the SAT attack, and the returned $|L|$-bit key value $K$ and the DIPs $P$ are saved, Line 32. The key $K$ is validated by checking individual bits with the corresponding equivalent fault representation.  The fault simulation is performed to identify the detected and redundant faults, Line 33. As we correctly determine the key ($K$), the test pattern must detect all the faults except the redundant ones that do not impact the functionality. As a result, undetected faults from the fault simulation must be redundant. These patterns in $P$ are recorded in the set $P_{A2}$, Line 34. Upon execution of the algorithm, four sets of redundant faults $RF_{A2}$, hard-to-detect fault $DF_{A2}$, test patterns $P_{A1}$ and $P_{A2}$ are reported back to the user, Line 35.

% \vspace{-10px}
\section{Result and Analysis} \label{sec:result}

% SAT program ends, no pattern returned
% emphasis on redundant faults 
In this section, we present the experimental results of our proposed SAT-based test pattern generation and redundant fault identification. The proposed miter construction with the SAT attack and test pattern generation are analyzed using ITC'99 benchmark circuits (\url{https://github.com/squillero/itc99-poli.}). %Although our approach targets combinational circuits, it is also applicable to any sequential circuit with scan chains. It can be converted to the equivalent combinational design by considering the I/O of the scan flip-flop as pseudo-primary input and pseudo-primary output. 
We use Synopsys 32nm SAED32 library to map the benchmark circuits to technology-dependent netlists, which are read in with TetraMAX II ATPG~\cite{SynopsysTetraMAX}. Any advanced technology nodes can also be applied to map the technology-independent \textit{bench} file with the standard cells in the library, as described in Section~\ref{sec:algorithm}. We first apply a commercial ATPG tool, Synopsys TetraMAX II, to generate test patterns for detecting all the \textit{sa0} and \textit{sa1} faults in a circuit. The tool reports test patterns, fault coverage, and undetected faults. We have not modified the existing test pattern generation process. To determine the hard-to-detect faults that are previously undetected and find the corresponding test vectors, we ($i$) replaced all undetected faults with their key-based equivalent gates, and ($ii$) apply the proposed Approaches 1 and 2 to obtain additional test patterns. Note that our proposed technique provides supplemental coverage in addition to the test results from TetraMax II.

Table~\ref{tab:itc} summarizes our findings with 25 combinational benchmarks from ITC'99. We excluded simple benchmark circuits, where the TetraMAX II detects all stuck-at faults. Any faults that have no pattern generated are labeled as undetected ones. All the undetected faults reported by TetraMAX II have been evaluated with the proposed logic locking-based fault representation and the SAT attack. We apply the proposed approaches for detecting these undetected faults. Column 2 shows the total gate count for each benchmark. The total number of stuck-at faults, $TF$, for each benchmark is recorded in Column 3. For TetraMAX II~\cite{SynopsysTetraMAX}, it includes faults under the following four categories, $PT$ (Possibly Detected), $UD$ (Undetectable), $AU$ (ATPG Untestable), and $ND$ (Not Detected) and shown in Columns 4, 5, 6, and 7 respectively. The total undetected fault count ($UF=PT+UD+AU+ND$) and fault coverage ($FC$) obtained from TetraMAX II are listed in Columns 8 and 9. Our proposed approach identifies these undetected faults as either redundant faults ($RF_P$) or hard-to-detect faults ($RF_P$), which are listed in Columns 10 and 11. Column 12 represents the total fault coverage ($FC_T$) after applying our proposed SAT-based test pattern generation in addition to TetraMAX II. We computed the total fault coverage for Column 12 using the following Equation:
\[
FC_T=\frac{DF_{TMAX}+DF_P+RF_P}{TF}\times 100, 
\]

\noindent where, $DF_{TMAX}$ is the number of detected faults from the TetraMAX II tool. For example, the $b20\_C$ benchmark has 118298 faults, out of which 684 faults are not detected by TetraMAX II. Our proposed approach detects 53 hard-to-detect faults and identifies the rest 631 faults as redundant. Note that many of the small circuits do not have any hard-to-detect faults (e.g., $b04\_C$, $b05\_C$, etc.), and all the undetected faults are redundant. For bigger benchmark circuits (e.g., $b19\_C$), we observe an increased number of both the hard-to-detect and redundant faults. Note that our approach can generate test patterns for all the hard-to-detect faults and identify all the redundant faults resulting in a perfect fault coverage of 100\%.    

\begin{table}[ht]
\caption{Hard-to-Detect Faults summary in ITC'99 Benchmarks.}\vspace{-10px}
\small
\begin{center}
\begin{tabular}{|M{1.45cm}|M{1cm}|M{1cm}|M{1cm}|M{1cm}|M{.75cm}|}
\hline
\multirow{2}{*}{\textbf{Benchmark}}   &    \multicolumn{4}{c|}{\textbf{Detected Fault by Category}}  & \textbf{Total}\\ \cline{2-5}
&     $\bm{D/PT}$ & $\bm{D/UD}$   & $\bm{D/AU}$  & $\bm{D/ND}$   & $\bm{DF_P}$ \\ \hline
b05\_opt\_C & 4/4   & 0/3       & 0/0   & 0/0     & 4    \\ \hline
b11\_C      & 0/0   & 0/161     & 0/0   & 5/5     & 5   \\ \hline
b15\_opt\_C & 3/19  & 0/1054    & 0/38  & 8/187   & 11 \\ \hline
b17\_opt\_C & 10/15 & 0/1153    & 0/300 & 16/1702 & 26 \\ \hline
b17\_C      & 0/0   & 0/3866    & 0/33  & 1/378   & 1  \\ \hline
b20\_C      & 35/35 & 0/631     & 18/18 & 0/0     & 53  \\ \hline
b21\_C      & 8/8   & 0/684     & 0/0   & 0/2     & 8   \\ \hline
b22\_C      & 32/33 & 0/795     & 29/29 & 0/0     & 61  \\ \hline
b18\_opt\_C & 0/0   & 0/479     & 0/17  & 14/62   & 14  \\ \hline
b18\_C      & 0/0   & 0/3658    & 0/96  & 4/10    & 4  \\ \hline
b19\_C      & 4/6   & 0/8236    & 0/191 & 95/126  & 99 \\ \hline
\end{tabular}
 \end{center}\label{tab:itc-detected}\vspace{-10px}
\end{table}

We identify these $DF_P$s from the four undetected fault categories reported by TetraMAX. Table~\ref{tab:itc-detected} shows the number of hard-to-detect faults from $PT,~UD,~AU$, and $ND$ categories. The second column represents the additional detected faults ($D$) from $PT$ and is presented as $D/PT$. Similarly, Columns 3, 4, and 5 show additional detected faults from $UD, AU$, and $ND$, respectively. We have detected a few faults from $PT, AU$, and $ND$ categories, except $UD$ categories. For example, 32 faults from $PT$ and 29 faults from $AU$ are detected for \textit{b22\_C} benchmark. Similarly, 95 out of 126 faults are detected from the $ND$ category for \textit{b19\_C} benchmark. However, we did not observe any detected faults from the $UD$ category, and they are all redundant. In summary, we found that some faults from all the other categories, except $UD$, are hard-to-detect while others are redundant.

\begin{table}[ht] %\vspace{-10px}
\caption{Comparison on number of test patterns on SAT detected faults between Approach 1 and Approach 2.\vspace{-10px}}
\small
\begin{center}
\begin{tabular}{|c|c|c|c|}
\hline
{\textbf{Benchmark}} & {\centering \textbf{Approach 1}}& {\centering \textbf{Approach 2}} & \textbf{Reduction}\\ \hline

b05\_opt\_C& 4 &  1 & 75.00\% \\ \hline
b11\_C& 5 &  2 & 60.00\%\\ \hline
b15\_opt\_C& 11 &  7 & 36.36\%\\ \hline
b17\_opt\_C& 26 &  12  & 53.85\%\\ \hline
b17\_C     & 1 &  1 &  0\%\\ \hline
b20\_C     & 53 &  25 & 52.83\%\\ \hline
b21\_C     & 8 &  3  & 62.50\%\\ \hline
b22\_C     & 61 & 26 &  57.38\% \\ \hline
b18\_opt\_C& 14 & 7  & 50.00\%\\ \hline
b18\_C     & 4 & 1 & 75.00\%\\ \hline
b19\_C     & 99 & 8 & 91.92\% \\\hline
% voter      & 1686 &  \\ \hline
% mem\_ctrl  & 4 & 1 \\ \hline
\end{tabular}
\end{center}\label{tab:approach2}
\vspace{-10px}
\end{table}
% \todo{Do the same like before}
For each benchmark, we combine all faults in the hard-to-detect fault set $DF_P$ to generate the optimized test set with the proposed Approach 2 and the SAT attack, presented in Section~\ref{sec:testpatterngen}. Table~\ref{tab:approach2} compares the number of test patterns required for Approach 1 and Approach 2. Columns 2 and 3 record the total test pattern count for Approach 1 and Approach 2 on $DF_P$, respectively. Column 3 represents the percentage decrease in the number of test patterns between Approach 2 and Approach 1. As shown in Table~\ref{tab:approach2}, we can see a significant reduction in the number of test patterns required to identify the faults, {with an average of 52.29\% fewer test vectors}. For example, the 61 hard-to-detect faults in $DF_P$ in \textit{b22\_C} benchmark need 61 test patterns with Approach 1 but only 26 test vectors using the proposed Approach 2. The fault simulation validated all the input vectors returned by the SAT attack. In addition, all the key bits in each locked circuit have been validated with the proposed miter construction for stuck-at faults, and they all match the expected key values. 

\begin{table*}[h]
\caption{Test Generation Time Summary for ITC'99 Benchmarks.}\vspace{-10px}
\small
\begin{center}
\begin{tabular}{|c|c|c|c|c|M{1.5cm}|c|M{1.5cm}|c|}
\hline
\multirow{2}{*}{\textbf{Benchmark}} & \multirow{2}{*}{\# $\bm{DF_P}$} & \multirow{2}{*}{\# $\bm{UF}$}  & \multicolumn{2}{c|}{\textbf{Approach 1 (s)}} & \multicolumn{2}{c|}{\textbf{Approach 2 (s)}} & \multicolumn{2}{c|}{\textbf{Improvement ($\times$)}}\\ \cline{4-9}
& && \multicolumn{1}{c|}{$\bm{DF_P}$}   & \multicolumn{1}{c|}{$\bm{DF_P+RF_P}$} & \multicolumn{1}{c|}{$\bm{DF_P}$}  & \multicolumn{1}{c|}{$\bm{DF_P+RF_P}$} & \multicolumn{1}{c|}{$\bm{DF_P}$}  & \multicolumn{1}{c|}{$\bm{DF_P+RF_P}$} \\ \hline
b04\_opt\_C & 0  & 5    & –           & 0.43        & –       & 0.07   & –     & 5.8        \\ \hline
b04\_C      & 0  & 50   & –           & 4.9         & –       & 0.11   & –     & 44.5       \\ \hline
b05\_opt\_C & 4  & 7    & 0.60        & 0.77        & 0.15    & 0.08   & 4     & 9.6        \\ \hline
b05\_C      & 0  & 1103 & –           & 576.4       & –       & 558.7  & –     & 1.0        \\ \hline
b07\_opt\_C & 0  & 1    & –           & 0.05        & –       & 0.05   & –     & 1.0        \\ \hline
b07\_C      & 0  & 6    & –           & 0.41        & –       & 0.08   & –     & 5.5        \\ \hline
b11\_opt\_C & 0  & 2    & –           & 0.20        & –       & 0.07   & –     & 2.8        \\ \hline
b11\_C      & 5  & 166  & 0.54        & 58.7        & 0.10    & 0.14   & 5.4   & 419.9      \\ \hline
b12\_opt\_C & 0  & 1    & –           & 0.10        & –       & 0.10   & –     & 1.0        \\ \hline
b13\_C      & 0  & 80   & –           & 2.3         & –       & 0.05   & –     & 42.7       \\ \hline
b14\_opt\_C & 0  & 319  & –           & 414.9       & –       & 11.6   & –     & 35.6       \\ \hline
b14\_C      & 0  & 38   & –           & 71.9        & –       & 30.9   & –     & 2.3        \\ \hline
b15\_opt\_C & 11 & 1298 & 62.4        & 4,122.9     & 9.9     & 170.4  & 6.3   & 24.2       \\ \hline
b15\_C      & 0  & 1518 & –           & 4,071.6     & –       & 10.9   & –     & 374.2      \\ \hline
b17\_opt\_C & 26 & 3170 & 457.8       & 25,967.4    & 22.5    & 4780.1 & 20.3  & 5.4        \\ \hline
b17\_C      & 1  & 4277 & 16.8        & 31,559.3    & 16.8    & 133.7  & 1     & 236.0      \\ \hline
b20\_opt\_C & 0  & 104  & –           & 334.8       & –       & 22.3   & –     & 15.0       \\ \hline
b20\_C      & 53 & 684  & 139.5       & 1,764.3     & 6.8     & 135.8  & 20.5  & 13.0       \\ \hline
b21\_opt\_C & 0  & 106  & –           & 284.7       & –       & 22.8   & –     & 12.5       \\ \hline
b21\_C      & 8  & 694  & 15.7        & 2,085.4     & 1.2     & 247.7  & 13.1  & 8.4        \\ \hline
b22\_opt\_C & 0  & 169  & –           & 731.7       & –       & 31.0   & –     & 23.6       \\ \hline
b22\_C      & 61 & 857  & 257.2       & 4,021.4     & 8.2     & 584.6  & 31.4  & 6.9        \\ \hline
b18\_opt\_C & 14 & 558  & 96,199.7    & 7,045,240.5 & 1,518.7 & 1471.1 & 63.3  & 4789.1     \\ \hline
b18\_C      & 4  & 3764 & 12,474.5    & \textit{timeout}     & 1,992.7 & 3740.7 & 6.3   & $>$1000    \\ \hline
b19\_C      & 99 & 8559 & 2,083,961.1 &  \textit{timeout}     & 8,728.8 & 4675.1 & 238.7 & $\gg$1000  \\ \hline
\end{tabular}
 \end{center}\label{tab:test-pattern-generation-time}\vspace{-15px}
\end{table*}

We run our proposed algorithms on a 20-core Intel Xeon CPU with 2.60 GHz and 64 GB RAM. The SAT program runs on a single thread in CentOS Linux 7 operating system. We only consider the SAT attack time as the preprocessing, such as technology-dependent netlists to technology-independent bench file conversion, processing of TetraMAX report, fault simulation, etc., can be performed in parallel. Table~\ref{tab:test-pattern-generation-time} shows the runtime for detecting both hard-to-detect faults $DF_P$ and the redundant faults $RF_P$ using our proposed Approaches 1 and 2. Note that we compare the time complexity for our proposed approaches only due to the --$(i)$ unavailability of programs for prior SAT-based approaches in the public domain, and $(ii)$ many optimizations performed over the years for commercial ATPG tools. As our server has 20 cores, a total of 40 threads that target 40 faults can be run in parallel. We set 5 days as the timeout for all undetected faults. Columns 2 and 3 are the fault count for $DF_P$ and total undetected faults ($UF$) from TetraMAX II, respectively. Note that $UF=DF_P+RF_P$ as we either detect all these faults or identify them as redundant. Columns 4 and 5 show the total test time for $DF_P$ and $DF_P+RF_P$ using our proposed Approach 1. The same is shown in Columns 6 and 7 for Approach 2. Columns 8 and 9 represent the test time improvement of Approach 2 over Approach 1. Please note that $\#DF_P$ fault count (Column 2) for b18\_C and b19\_C are computed from fault simulation with the test patterns obtained from the SAT attack with $DF_P+RF_P$ and Approach 2 (Column 7), where a timeout of 5 days for a single fault is observed, marked as \textit{timeout} in Column 5. For example, Approach 1 takes 0.54s, whereas Approach 2 requires only 0.10s to detect 5 hard-to-detect faults for the b11\_C benchmark circuit. However, the gain for Approach 2 becomes significant for larger benchmark circuits. For b19\_C, Approach 1 takes 2,083,961.1s to detect 99 faults, which is much larger than Approach 2, 8728.8s. The improvement in test time is $2,083,961.1/8,728.8=238.7$ times. \textit{This signifies the fact that detecting 99 $DF_P$s together is easier than detecting a single fault. Interestingly, generating test patterns even for detecting and identifying 8559 faults as $DF_P$ and $RF_P$ takes less time than detecting 99 $DF_P$s.} We observe that the overall gain becomes significant when there are an increased number of faults to be grouped during test pattern generation, and the number of conflicts typically increased with the benchmark size. The reason for this increase in the ratio is due to the reduction of conflicts resulting from the initialization of the multiple key bits in the SAT-attack tool, as shown in Figure~\ref{fig:approach1}. 

\vspace{-10px}
\section{Conclusion} \label{sec:conclusion}
In this paper, we presented how the widely explored SAT attack on logic locking can be used to identify redundant faults and generate test patterns for hard-to-detect stuck-at faults. We first present the miter construction of stuck-at faults to a key-dependent locked circuit so that the powerful SAT tool can be used. This ensures that the input patterns used to break our logic locking technique can be applied to detect the stuck-at faults. Since the SAT-based attack effectively breaks multiple logic locking schemes, we exploit it to generate test vectors for stuck-at faults with the corresponding locked circuits. If faults are observable at the primary output, the distinguishing input patterns returned from the SAT attack can expose them. On the other hand, if any faults are redundant, no distinguishing input can be found from the SAT attack, and the program finishes directly with the UNSAT conclusion from the SAT solver. By applying our proposed approach, we were able to identify any redundant faults from the undetected stuck-at faults reported by the ATPG tool or obtain the necessary test patterns for those non-redundant hard-to-detect faults. Our test pattern generation approach can also be optimized for a reduced pattern set by grouping multiple faults into a single locked circuit. In the future, we plan to explore test pattern generation without a commercial ATPG tool. In manufacturing tests, test time per chip plays an important role after chip fabrication. Keeping a low number of test patterns becomes one of the prime objectives for VLSI testing with less time in addition to achieving the desired fault coverage. We plan to apply random patterns to the fault simulator to detect a majority of the easy-to-detect faults and then apply the proposed Approach 2. In addition, we plan to study test pattern compaction to reduce pattern count.

\vspace{-15px}
\section*{ACKNOWLEDGEMENT}
This work was supported by the National Science Foundation under Grant Number CNS-1755733. %Any opinions, findings, conclusions, or recommendations expressed in this material are those of the authors and do not necessarily reflect the views of the National Science Foundation.

% \balance
\bibliographystyle{IEEEtran} %{alpha}
% Generated by IEEEtran.bst, version: 1.14 (2015/08/26)

\vspace{-30px}
\begin{IEEEbiography}[{\includegraphics[width=1\linewidth]{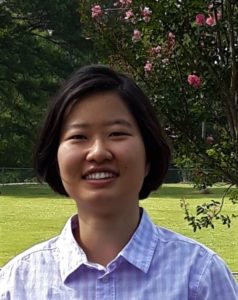}}]{Yadi Zhong (S'20)} is currently pursuing her Ph.D. in Computer Engineering from the Department of Electrical and Computer Engineering, Auburn University, AL, USA. She received her B.E. degree from the same university in 2020. Her research interests are logic locking, fault injection and hardware security, and post-quantum cryptography. She received the Best Paper award at IEEE Physical Assurance and Inspection of Electronics (PAINE'22). She also led a student team that received several awards including 1st place in Hack@CHES 2021 and 2nd place in Hack@SEC 2021. She is a student volunteer for HOST 2022. She was also the recipient of the Auburn University Presidential Graduate Research Fellowships in 2020. She is a student member of the IEEE.
\end{IEEEbiography}

\vspace{-30px}
\begin{IEEEbiography}[{\includegraphics[width=1in,height=1.25in,clip,keepaspectratio]{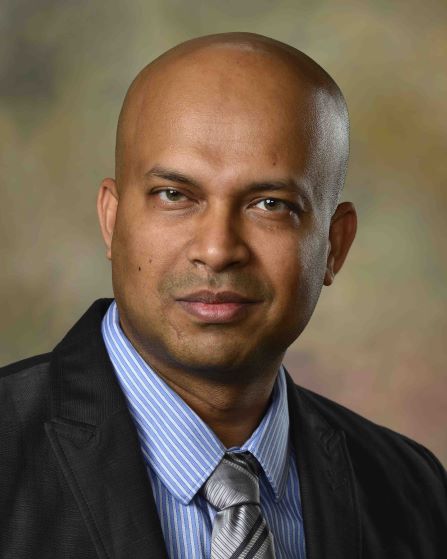}}]{Ujjwal Guin (S'10--M'16--SM'22)} received his PhD degree from the Electrical and Computer Engineering Department, University of Connecticut, in 2016. He is currently an Assistant Professor in the Electrical and Computer Engineering (ECE) Department of Auburn University, Auburn, AL, USA. He received his B.E. degree from the Department of Electronics and Telecommunication Engineering, Bengal Engineering and Science University, India, in 2004 and his M.S. degree from the ECE Department, Temple University, Philadelphia, PA, USA, in 2010. Dr. Guin's current research interests include hardware security, blockchain, and VLSI design \& test. He has authored several journals and refereed conference papers. He serves on organizing committees of HOST, VTS, ITC-India, and PAINE. He also serves on technical program committees in several reputed conferences, such as DAC, HOST, ITC, VTS, PAINE, ICCD, GLSVLSI, ISVLSI, and Blockchain. He is a senior member of IEEE.
\end{IEEEbiography}

\end{document}